\documentclass[reprint,amsmath,amssymb,amsfonts,aps,prb,superscriptaddress]{revtex4-2}
\usepackage{graphicx}
\usepackage{bm}
\usepackage{epsfig}
\usepackage{epstopdf}
\usepackage{xcolor}
\usepackage{hyperref}
\hypersetup{colorlinks=true,allcolors=blue}
\usepackage{natbib}
\usepackage{mathtools}
\usepackage{siunitx}
\usepackage{soul,xcolor}
\usepackage{amsmath,amssymb}

\setstcolor{red}
\DeclareUnicodeCharacter{2212}{-}
\DeclarePairedDelimiterXPP\BigOSI[2]%
  {\mathcal{O}}{(}{)}{}%
  {\SI{#1}{#2}}

\begin{document}
\title{ Strong Disorder Renormalization Group Method
for Bond Disordered Antiferromagnetic Quantum Spin Chains with Long Range Interactions: Ground State Properties}

\author{S. Kettemann}
\email[]{skettemann@constructor.university}
\affiliation{Department of Physics and Earth Sciences and 
Department of Computer Science, Constructor University,  Campus Ring 1, 28759 Bremen, Germany}

\begin{abstract} 
We 
  introduce and implement  a reformulation of 
 the strong disorder renormalization group method in real space, 
 well suited
 to study
 bond disordered antiferromagnetic power law  coupled quantum spin chains. 
We  
 derive  the Master equations for the 
  distribution function of pair distances $\tilde{r}$.
 First, we  apply it to 
 a short range  coupled spin chain,    keeping only interactions for  adjacent spins.  We confirm  that it is solved by  the infinite randomness fixed point distribution. 
Then,  we solve  the 
  Master equation 
  for the power law long  range interaction between all spins for any anisotropy 
   ranging from the XX-limit to the isotropic Heisenberg limit, corresponding to a 
   tight binding chain of disordered 
   long range interacting Fermions with long range hopping. 
We  thereby  show that  the distribution function of couplings $J$ at renormalization scale $\Omega$
flows to the 
    strong disorder fixed point distribution  with small  corrections 
    at $\tilde{r} > \rho,$ which 
      depend
      on power exponent $\alpha$ and coupling anisotropy $\gamma.$
       As a consequence, the low temperature 
      magnetic susceptibility  diverges
       with an  anomalous power law.
The distribution of singlet lengths  $l$ is found to 
 decay as $l^{-2}$. The entanglement entropy of a subsystem of length $n$
increases  in the ground state logarithmically for all $\alpha$ and $\gamma$. 
 After a global quantum quench the entanglement entropy increases with time logarithmically
  as  $S(t) \sim   \ln(t)/(2\alpha)$. 
\end{abstract} 

\maketitle

\section{Introduction}
 Disordered quantum spin  systems 
    with  long range interactions
 govern  the  properties of a wide range of    materials, including   
   doped semiconductors  with randomly placed magnetic dopants \cite{anderson58,loehneysen,Bhatt2021,Kettemann2023},
   metals with magnetic impurities\cite{Kettemann2024} and
   glasses whose low temperature properties are dominated by 
    the dynamics of 
   tunneling systems, coupled by dipolar and elastic interactions\cite{reviewtls,Yu,Bilmes2020}. 
   Recently, it became possible to study 
     disordered spin ensembles at a diamond surface, which    can be  probed
         with  single nitrogen-vacancy centers in diamond
               \cite{Lukin2022,Davis2022}.
 Tunable interactions have been realized  between atoms trapped near photonic crystals
 \cite{douglas},
 \cite{grass} and by coupling  Rydberg states with opposite parity \cite{Signoles2021,Brow2020,Franz2022}. 
 Trapped ions with power-law interactions, decaying as $1/r^\alpha$, with tunable $0<\alpha<1.5$  have  been realized\cite{Islam2013,Richerme2014,Jurc2014}.
   However, it remains a challenge to derive thermodynamic and dynamic  properties 
    of  such  systems. 
   The long range interactions demand  the study
   of large system sizes and  the disorder necessitates to  
   obtain a large number of ensembles of disorder realizations. This  limits the 
   potential
   of 
   numerical calculations to tackle such problems.

 The  strong disorder renormalization group  (SDRG) method
  has been developed and successfully applied  to study
disordered   quantum spin chain models, allowing the 
    derivation of  their thermodynamic and dynamic properties\cite{bhattlee81,bhattlee82,fisher94,fisher95,monthus,Igloi2018}.
This 
 has lead to the  discovery of the infinite randomness fixed point (IRFP) of short range coupled disordered spin chain models\cite{bhattlee81,bhattlee82,fisher94,fisher95,monthus,Igloi2018}.
    At the IRFP the ground state   is 
    composed of  randomly placed 
    pairs of spins in 
     singlet states, the {\it random singlet phase}.  
    The SDRG method procedure has been  
    originally applied to 
    short range,  bond disordered
           antiferromagnetic spin $S=1/2$-chains\cite{bhattlee81,bhattlee82,fisher94,fisher95} with an  
          initial  distribution of couplings $J,$
           $P(J,\Omega_0),$ 
           where  
           $\Omega_0$ is  the  largest energy scale in the spin chain. 
 Identifying 
           the  pair of spins   $(i,j)$ which 
          are coupled by the largest coupling  $J_{ij} = \Omega$ defines the renormalization group (RG) scale $\Omega$.
For antiferromagnetic coupling that pair 
of spins is bound into  a singlet state,
  inducing  a coupling between its
 adjacent spins $J_{kl},$  which is  a  function of  the removed couplings $J_{li}$ and $J_{jk}$ and defines
   the RG rule.
    As this new coupling is generated,  the 
     distribution of couplings  is modified to   $P(J,\Omega).$
     Repeating this procedure   until all spins are paired in singlets, one arrives at a product state of singlets with a  coupling distribution 
     $P(J, \Omega \rightarrow 0),$  
given by  
\begin{equation} \label{pj}
  P(J,\Omega) =\frac{1}{\Omega \Gamma_{\Omega}} \left( \frac{\Omega}{J}\right)^{1-1/\Gamma_{\Omega}} \theta(\Omega-J),
      \end{equation}        
      whose  width  $\Gamma_{\Omega} = \ln \Omega_0/\Omega$ diverges as $\Omega\rightarrow 0$.
       Here, $\theta(x)$  is the unit step function,
   $\theta(x>0) =1$, and   $\theta(x<0) =0.$
      This
      distribution holds also for  other   short range random  quantum  spin chains\cite{monthus}.
         Recently, 
        the  SDRG  method
          was extended
          to  disordered  spin   $S=1/2-$chains
         with antiferromagnetic  long range 
          interactions.
         All couplings are found to be renormalized according to 
         new SDRG rules. Implementing these rules
         for XX-coupled spin chains 
          the 
           ground state  was shown to be  
           a product state of random singlets 
           with a fixed point distribution of their couplings
          given by Eq. (\ref{pj})  but with finite width\cite{Moure2015,Moure2018,Mohdeb2020},
              $
              \Gamma (\Omega) \rightarrow 2 \alpha.$ 
           This was confirmed by 
          numerical exact diagonalization  and   
            DMRG methods\cite{Mohdeb2020}.  A similar  fixed point distribution  was  found for long range coupled 
           transverse field Ising chains
           \cite{Juhasz2014,Juhasz2016}.
            The SDRG method was 
            extended to   SDRG-X 
for excited  states 
\cite{Pekker2014}
and 
 to  the SDRG-t method to study entanglement dynamics 
 \cite{Vosk2013,Vosk2014,Igloi2012}
 These methods were recently applied to 
             excited states \cite{Mohdeb2022} and  
entanglement  dynamics  after global quantum quenches
\cite{Mohdeb2023} in long range coupled spin chains.         
It remains to derive properties of long range bond disordered
quantum  Heisenberg spin chains,  corresponding to chains of   interacting  Fermions with disordered  long range hopping and interactions\cite{Moure2015}.
 To this end, we introduce  a novel representation of the SDRG method in real space.  

\begin{figure}
    \includegraphics[scale=0.15]{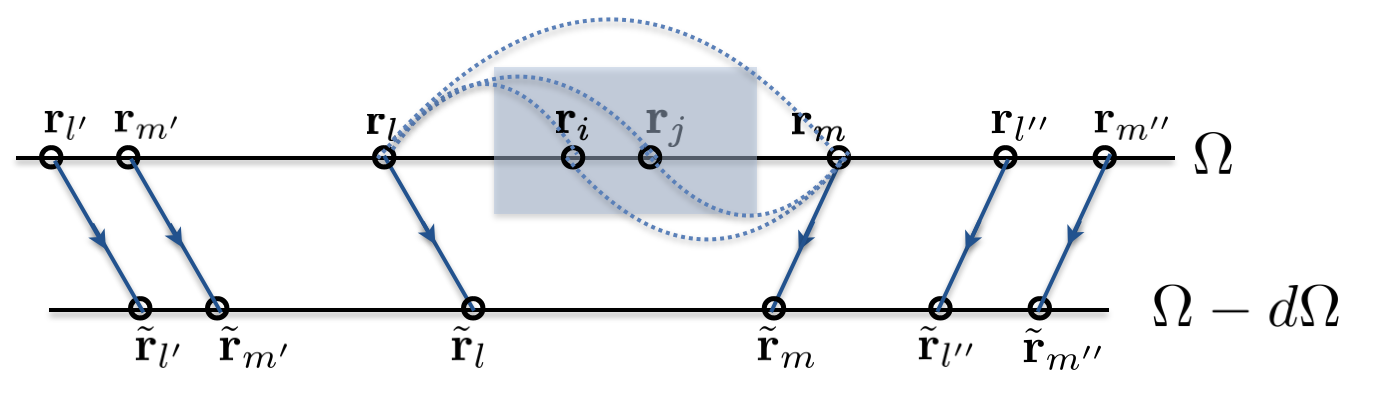}
\vspace*{-.3cm}\caption{SDRG step in real space for a chain of randomly placed spins  (circles): The decimation of the strongest coupled spin pair $(i,j)$ (shaded), 
whose coupling defines the  RG scale $\Omega,$ 
is followed by the renormalization of all positions of spins,  ${\bf r}_{l} \rightarrow \tilde{{\bf r}}_{l}$ as the RG scale  $\Omega - d\Omega$ is reduced.} 
\vspace*{-.5cm}
    \label{RG}
\end{figure}

\begin{it}
Model.\end{it}— 
We study long range antiferromagnetically  coupled spin chains with  $N$ 
$S=1/2-$spins, randomly placed at positions ${\bf r}_i$ 
on a line
of length $L,$ Fig. \ref{RG}
\begin{equation}\label{H}
H=\sum_{i<j} \left[ J^x_{ ij}\left(S_{i}^{x}\,S_{j}^{x}+S_{i}^{y}\,S_{j}^{y}\right) + 
 J^z_{ ij} S_{i}^{z}\,S_{j}^{z} 
\right].
\end{equation}
  The couplings between spins at sites $i,j$ are antiferromagnetic and long-ranged, decaying with a power law  with distance $r_{ij} = |{\bf r}_i-{\bf r}_j|$ with  exponent $\alpha$, 
\begin{equation} \label{jcutoff}
J^{\kappa}_{ij} =
J^{\kappa}_0\left|({\bf r}_i-{\bf r}_j)/a\right |^{-\alpha},
\end{equation}
with $\kappa \in \{x,z \}.$
 The anisotropy is parametrized by 
 $\gamma = J^z/J^x = J_0^z/J_0^x .$ We consider open boundary conditions.
 We review in appendix A the 
  mapping of  Eq. (\ref{H}) to a tight binding Hamiltonian of  Fermions 
  by the Jordan-Wigner transformation. There, it is seen that the long range hoppings cause
  phase correlations, 
  which 
 make the Hamiltonian    a  challenging correlation problem
       even  in the absence of direct interactions, $J^z_{ij}=0.$
       A finite coupling $J^z_{ij} \neq 0$ introduces  the direct  interaction between fermions at different sites $i,j,$
         making it an interacting many body problem. 
 
  Applying  the  SDRG
             procedure 
             to the anisotropic spin chain 
             Eq. (\ref{H}), 
            we identify  the strongest coupling  
              $J_{ij} = \Omega,$
               highlighted
               in the  example of randomly placed spins
 in Fig.  \ref{RG}.
 Thereby,  the 
               spin pair $(i,j)$ is forced 
              into a singlet state,   
            and the 
             couplings between all remaining pairs of  spins $(l,m)$ are renormalized to\cite{Moure2015}
              \begin{eqnarray} \label{jeff}
               \tilde{J}^{x}_{lm} &=&   J_{lm}^x - \frac{(J^x_{li}-J^x_{lj})(J^x_{im}-J^x_{jm})}{J^x_{ij}+J^z_{ij}}, 
               \nonumber \\
               \tilde{J}_{lm}^{z} &=&   J^z_{lm} - \frac{(J^z_{li}-J^z_{lj})(J^z_{im}-J^z_{jm})}{2 J^x_{ij}},
              \end{eqnarray}
             which
              depend on the initial 
              coupling between  them, as well as on the couplings between 
               removed spins $(i,j),$ and 
               the spins   $(l,m),$ 
               indicated by blue lines
                 in Fig.  \ref{RG}.

\begin{figure}
    \includegraphics[scale=0.2]{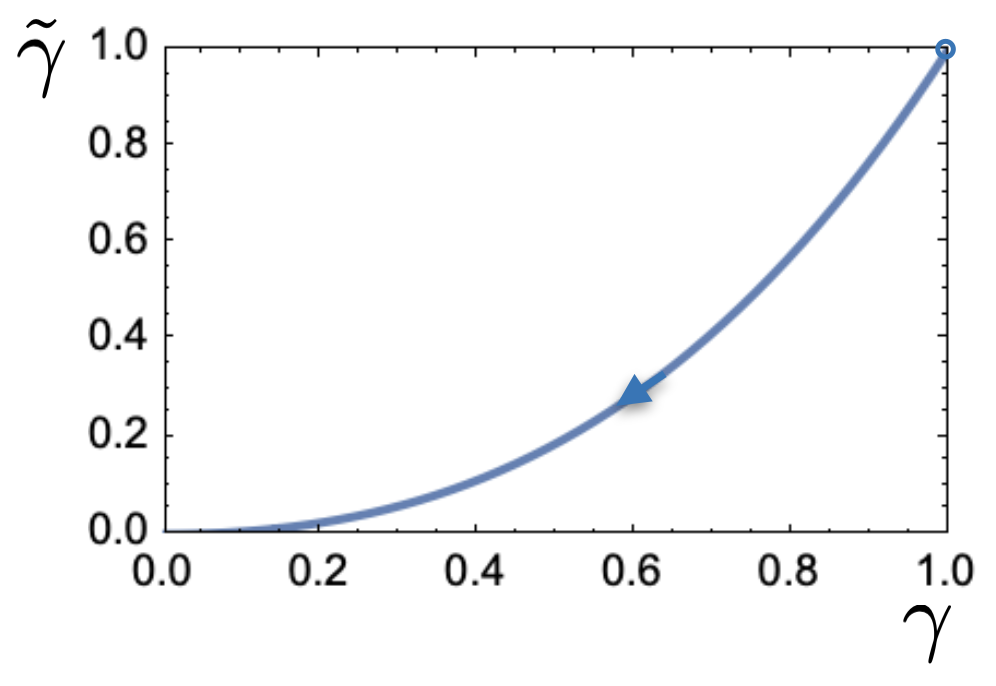}
\vspace*{-.3cm}\caption{Renormalized anisotropy  $\tilde{\gamma}$ as function of bare anisotropy $\gamma.$} 
\vspace*{-.5cm}
    \label{anis}
\end{figure}

 These renormalized  couplings $\tilde{J}$
can be  recast in terms of renormalized distances $\tilde{r},$
as $\tilde{J}^x_{lm}=J^x_0/\tilde{r}_{lm}^{\alpha}$
and $\tilde{J}^z_{lm}=\tilde{J}^z_0/\tilde{r}_{lm}^{\alpha}.$
The renormalization of the  anisotropy $\gamma$ is accounted for by the 
renormalized z-component $\tilde{J}^z_0$. 
The RG rules Eq. (\ref{jeff}) can  be 
reformulated  by insertion of  Eq. (\ref{jcutoff}) in terms of renormalized distances $r_{lm}=r \rightarrow \tilde{r}$,  Fig. \ref{RG}, yielding 
at  RG length scale $\rho = (\Omega/J_0)^{-1/\alpha},$ 
     \begin{eqnarray} \label{reff}
                &&\tilde{r}^{-\alpha} = r^{-\alpha} \times
                \nonumber \\
&&  \left( 1 +\frac{1}{1+\gamma} \left(\frac{r \rho}{r_{lj}r_{jm}}\right)^{\alpha}
                ((\frac{r_{lj}}{r_{li}})^{\alpha}-1)(1-(\frac{r_{jm}}{r_{im}})^{\alpha})\right).
              \end{eqnarray}
              For given  
               $\rho,$ the distance between the 
              removed spins,  
 there are constraints on the other
 five distances appearing in the 
 RG rule Eq. (\ref{reff}), which
 depend only on  two independently variable distances $R_L = r_{li}$ and $R_R = r_{jm},$
 see Fig. \ref{RG}.
The anisotropy parameter is found to be   renormalized to $\tilde{\gamma} = \gamma^2(1+\gamma)/2, $  which is  the same
renormalization rule as for the anisotropic  short ranged model\cite{monthus}.
$\gamma =1$ is a fixed point. 
For any smaller  initial anisotropy parameter  $0<\gamma <1,$
 the anisotropy  flows to the XX-spin fixed point,  as shown in Fig.  \ref{anis}.

The RG rule Eq. (\ref{reff}) is valid for any pair of 
spins $(l,m).$ It is convenient to be implemented numerically. 
  In the following, we rather aim to  
  derive the distribution function $P(\tilde{J},\Omega)$ analytically. 
In the representation of distances $\tilde{r}$ this corresponds to the 
 distribution function  
 $P(\tilde{r},\Omega) = (\alpha \tilde{J}/\tilde{r}) P(\tilde{J},\Omega)|_{\tilde{J}=\Omega_0 \tilde{r}^{-\alpha} }.$

 \section{
Nearest Neighbor  Coupling}

\begin{figure}
    \includegraphics[scale=0.35]{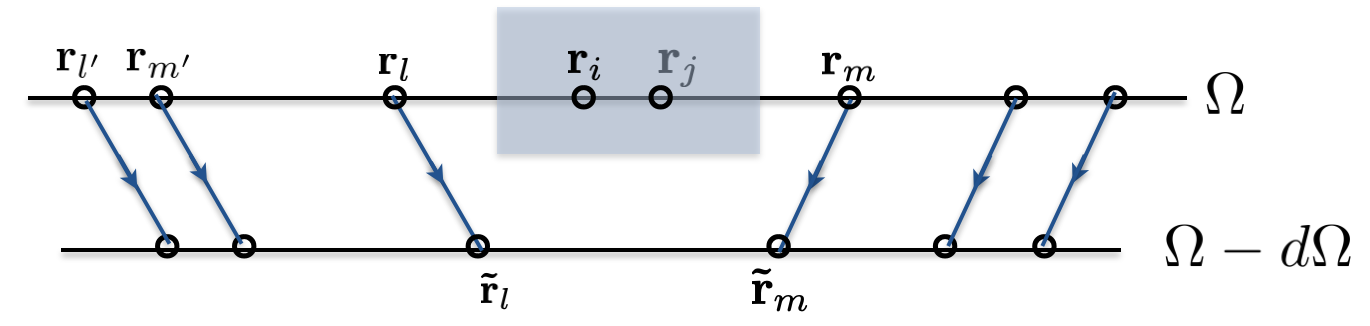}
\vspace*{-.5cm}\caption{Strong disorder RG step
for bond disordered short range coupled spin chains: Decimation of strongest coupled spin pair $(i,j)$,
highlighted by the shaded area, 
whose coupling defines the  RG scale $\Omega.$ It
is followed by renormalization of the positions of spins,  ${\bf r}_{l} \rightarrow \tilde{{\bf r}}_{l}$ and a reduction of the RG scale to $\Omega - d\Omega.$ } 
    \label{RGnn}
\end{figure}

Let us   first apply   the  SDRG
             procedure 
             to the anisotropic spin chain of randomly placed spins on a chain, 
   with power law coupling, Eq. (\ref{H}),   
              keeping only the coupling between adjacent spins. 
     Setting   in Eq. (\ref{jeff}) all couplings between non-adjacent pairs to zero, 
     we find the RG rules for the newly generated  couplings between the spin pair at sites $l,m,$
   see  Fig. \ref{RGnn}
           \cite{Moure2015}
              \begin{eqnarray} \label{jeffnn}
               \tilde{J}^{x}_{lm} &=&   + \frac{J^x_{li}J^x_{jm}}{J^x_{ij}+J^z_{ij}}, 
               \nonumber \\
               \tilde{J}_{lm}^{z} &=&  +  \frac{J^z_{li}J^z_{jm}}{2 J^x_{ij}}.
              \end{eqnarray}

Next we reformulate  these  RG rules in terms of the   renormalized distance $r_{lm}=r
\rightarrow \tilde{r},$
as sketched  in Fig. \ref{RGnn}. We note that 
 all other distances 
between adjacent spins remain unchanged.
At the RG length scale $\rho = (\Omega/J_0)^{-1/\alpha}$
the renormalized distance is thus 
     \begin{eqnarray} \label{reffnn}
                &&\tilde{r} =g_{\gamma}
               \frac{r_{li} r_{jm}}{\rho},
              \end{eqnarray}
              with anisotropy factor $g_{\gamma} = (1+\gamma)^{1/\alpha}.$

The 
 distribution function of distances  $P(\tilde{r},\Omega)$  is governed by a 
 Master equation, which can be  derived as follows: 
when the singlet between 
spin pair  $(i,j)$ is formed
 at RG scale $\Omega = J_{ij},$
 the  couplings between  spin pairs
  $(l,i)$ and $(j,m)$ are taken away
   while a coupling between 
    spin pair $(l,m)$ is newly created with 
    renormalized coupling $\tilde{J}_{lm}.$
     In the representation of distances 
      this corresponds to take 
      the edges between spin pairs
  $(l,i)$ and $(j,m)$ 
  with distances  $r_{l,i}=R_L$ and $r_{j,m}=R_R$
  away and to create an edge between spin pair 
  $(l,m)$ with renormalized distance $\tilde{r}_{lm}=\tilde{r},$
  as shown    in Fig.  \ref{RGnn}, 
replacing the bare distance 
${r}_{lm}={r}.$ 
Following the argumentation given in Ref. \cite{monthus}
 and adopting it to the distribution function of distances $P(r,\Omega),$ 
the distribution at RG scale $\Omega-d\Omega)$ lowered by  an infinitesimal amount  $d\Omega$
is related to the distribution function at RG scale $\Omega$ by 
\begin{eqnarray} \label{prnn}
 &&P(\tilde{r},\Omega-d\Omega) =
 ( P(r,\Omega)+d\Omega P(\Omega,\Omega)  \times
 \nonumber \\
 && \int_{\rho}^{\infty} dR_L  
 \int_{\rho}^{\infty}
 dR_R P(R_L,R_R,\Omega)
 \nonumber \\
 && 
 \left(
 \delta (\tilde{r}- g_{\gamma} \frac{R_L R_R}{\rho}) -  \delta (\tilde{r}-R_L )-\delta (\tilde{r}-R_R)\right) ) \nonumber \\
 && \frac{1}{1-2d\Omega  P(\Omega,\Omega)}.
      \end{eqnarray}  
Here, the second term 
       on the right hand side of Eq. (\ref{prnn})
       accounts for 
        the addition of a  renormalized bond at distance $\tilde{r}.$
         This occurs with   probability $d\Omega  P(\Omega,\Omega,$
     the probability to add a bond   
      in the RG step  of width $d\Omega$. 
    The following two  terms take into account the removal of the two bonds with distance $R_L$ and $R_R,$ respectively.
     These terms are integrated over all possible distances $R_L,$ $R_R$ exceeding $\rho,$ which is by definition 
      the smallest distance at RG step $\Omega.$
    In order to normalize the distribution function,  we need  to divide 
         the right side of Eq.   (\ref{prnn})
        by $1-2d\Omega  P(\Omega,\Omega$,
         the probability that bonds are not removed during  the RG step $d\Omega.$
Assuming that 
$R_L, R_R$ are independently distributed, the 
joint distribution function  $P(R_L,R_R,\Omega)_c$,
 can be factorized, into marginal distributions of $R_L$ and $R_R,$
  yielding
     $P(R_L,R_R,\Omega) =  P(R_L,\Omega) P(R_R,\Omega)$.
      
Next, performing the integrals over the last two delta functions and using the normalization  condition $\int_\rho^{\infty} d r P(r,\Omega) = 1,$
we find in the limit  $d\Omega \rightarrow 0$
the    Master equation 
for the short ranged model 
\begin{eqnarray} \label{mnn}
 &&-\frac{d}{d\Omega}  P(\tilde{r},\Omega) =
  P(\Omega,\Omega) \int_{\rho}^{\infty} dR_L  
 \int_{\rho}^{\infty}
 dR_R 
 \nonumber \\
 && 
 P(R_L,\Omega) P(R_R,\Omega) 
 \delta (\tilde{r}- g_{\gamma} \frac{R_L R_R}{\rho}).
      \end{eqnarray}  
Inserting the Ansatz 
\begin{equation} \label{ansatz}
P(\tilde{r},\Omega) =  \frac{c(\Omega)}{\tilde{r}} (\frac{\rho}{\tilde{r}})^{c(\Omega)}   ~ \theta(\tilde{r}/\rho-1),
\end{equation}
 where $\theta(x)$  is the unit step function,
   $\theta(x>0) =1$, and   $\theta(x<0) =0,$
   we find  
that the requirement that is   a solution of 
Eq. (\ref{mnn}), 
fixes   $c(\Omega) = \alpha/\Gamma_{\Omega},$
with $\Gamma_{\Omega} = \ln (\Omega_0/\Omega).$
and furthermore requires that 
$\ln (\rho / \tilde{r}) = g_{\gamma}^c (\ln (\rho / \tilde{r}) +\ln g_{\gamma}).$
Inserting $g_{\gamma}=(1+\gamma)^{1/\alpha}$
this condition becomes 
   $\ln (\rho / \tilde{r}) = (1+\gamma)^{1/\Gamma_{\Omega}} (\ln (\rho / \tilde{r}) + (1/\alpha)\ln (1+\gamma)),$
   with anisotropy $0\le \gamma\le 1.$
Thus, 
we find that in the limit of  $\Omega \rightarrow 0 \equiv \rho \rightarrow \infty \equiv \Gamma_{\Omega} \rightarrow \infty,$
the condition is fulfilled, confirming that the Ansatz is a solution of the Master equation for any anisotropy $0<\gamma<1$.  
   Transforming back to the couplings
  $ P(\tilde{J},\Omega) = \tilde{r}/(\alpha \tilde{J})  P(\tilde{r},\Omega)|,$
  we find the distribution of couplings 
  \begin{equation} \label{psdrg}
  P(\tilde{J},\Omega) = 
  \frac{1}{\Omega \Gamma_{\Omega}}(\frac{\Omega}{\tilde{J}})^{1-1/\Gamma_{\Omega}} 
  \theta(\Omega/\tilde{J}),
  \end{equation}
  which is the infinite randomness
   fixed point distribution function 
 with width  $\Gamma_{\Omega} = \ln (\Omega_0/\Omega),$
 diverging to infinity for $\Omega \rightarrow 0.$
We confirm that it is independent of anisotropy $\gamma,$ as was found previously in Ref. \cite{fisher94}.

\section{
Long Range Coupling}

\begin{figure}
    \includegraphics[scale=0.4]{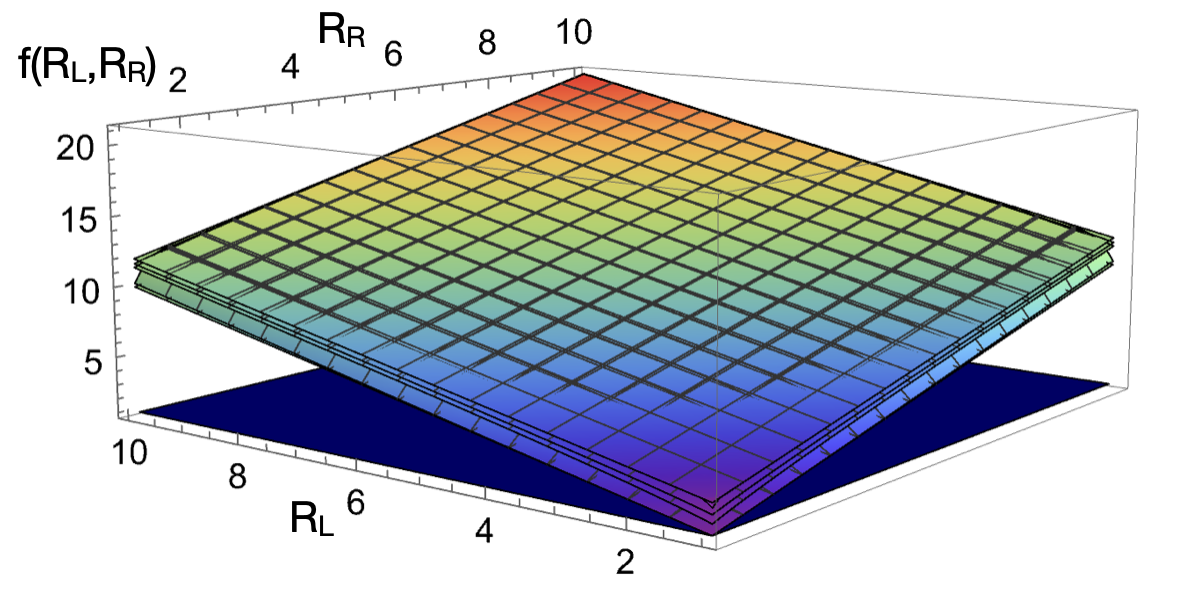}
\vspace*{-.5cm}
\caption{The renormalization 
function Eq. (\ref{rgelr}) as function of distances $R_L,R_R$
in units of $\rho,$
for $\alpha = 100,50,10,5,1, 0.5$ from the bottom up, 
as bounded by $r-2\rho= R_L+R_R-\rho$ from below 
and  $r=R_L+R_R+\rho$ from above.  } 
\label{ff}
\end{figure}

With long range coupling, we derive 
the  Master equation for 
the 
 distribution function   of distances
 $P(\tilde{r},\Omega)$ 
 in Appendix B, where it is found to be 
 given by 
      \begin{eqnarray} \label{mlfinalc}
 &&- \frac{d}{d \Omega} 
 P(\tilde{r},\Omega) =
P(\Omega,\Omega) \left( P(\tilde{r},\Omega) + C(\tilde{r},\Omega)  \right),
      \end{eqnarray}  
      where  we defined the function 
\begin{eqnarray} \label{c}
 &&
  C(\tilde{r},\Omega) =
 \int_{\rho}^{\infty}
 dR_L \int_{\rho}^{\infty}
 dR_R P(R_L,\Omega) P(R_R,\Omega) \times 
 \nonumber \\
 &&
 ( \delta (\tilde{r}- f(R_L,R_R,\rho))
 - \delta (\tilde{r}-(R_L+\rho+R_R)).
      \end{eqnarray}  
The renormalization function $f(R_L,R_R,\rho),$  is given in Appendix B, Eq.  (\ref{rgelr}).
It is a   highly nonlinear function 
 of $R_L,R_R$, as seen in Fig. \ref{ff}, where we plot it as  function  of $R_L,R_R$  for 
    various  values of $\alpha$.
We see there   that it is for all $\alpha$ bounded 
by 
$r-2\rho  < f(R_L,R_R,\rho) < r,$  where the upper limit 
corresponds to  the unrenormalized distance $r= R_L + \rho +R_R$. 
     The renormalization function $f$  drops towards
       $r-2\rho$  for small values of 
       $R_L,R_R,$
 while for larger values it approaches the bare value $r.$    
The crossover between these limits occurs  at 
  the line defined by  $\rho (R_L+R_R+\rho)/(R_L R_R) =1+\gamma.$

\subsection{  Bare Long Range  Couplings: Strong Disorder Fixed Point  Distribution}

 Including all long range couplings, but 
  neglecting the renormalization, by setting the renormalized distance to the bare distance
 $\tilde{r} = r,$ 
 or $  C(\tilde{r},\Omega)=0$ in 
 the  Master equation Eq. (\ref{mlfinalc}), it  simplifies to 
\begin{eqnarray} \label{pr}
 &&-\frac{d}{d\Omega}  P^0(\tilde{r},\Omega) =
  P^0(\Omega,\Omega)  P^0(\tilde{r},\Omega).
      \end{eqnarray}  
     This has the solution  
     \begin{eqnarray} \label{sdrgpr}
     P^0(\tilde{r},\Omega) = \rho^{1/2}/(2 \tilde{r}^{3/2})
     \theta (\tilde{r}-\rho).
      \end{eqnarray}
    Transforming back  to the distribution of couplings
  $ P^0(\tilde{J},\Omega),$  we recover the SDRG fixed point distribution Eq. (\ref{psdrg}) 
 with finite width $\Gamma = 2 \alpha.$ Thus, we find that   the  random 
 bare couplings $J = \Omega_0 r^{-\alpha}$ 
 result in  the SDRG distribution with finite width, not only 
 for $\gamma=0$, as  derived in 
 Ref. \cite{Mohdeb2020}, but 
 for the Heisenberg model Eq. (\ref{H}) with  any anisotropy $0 \le \gamma \le 1$. 

\subsection{ Distribution of  Renormalized Long Range Couplings}

\begin{figure}
    \includegraphics[scale=0.22]{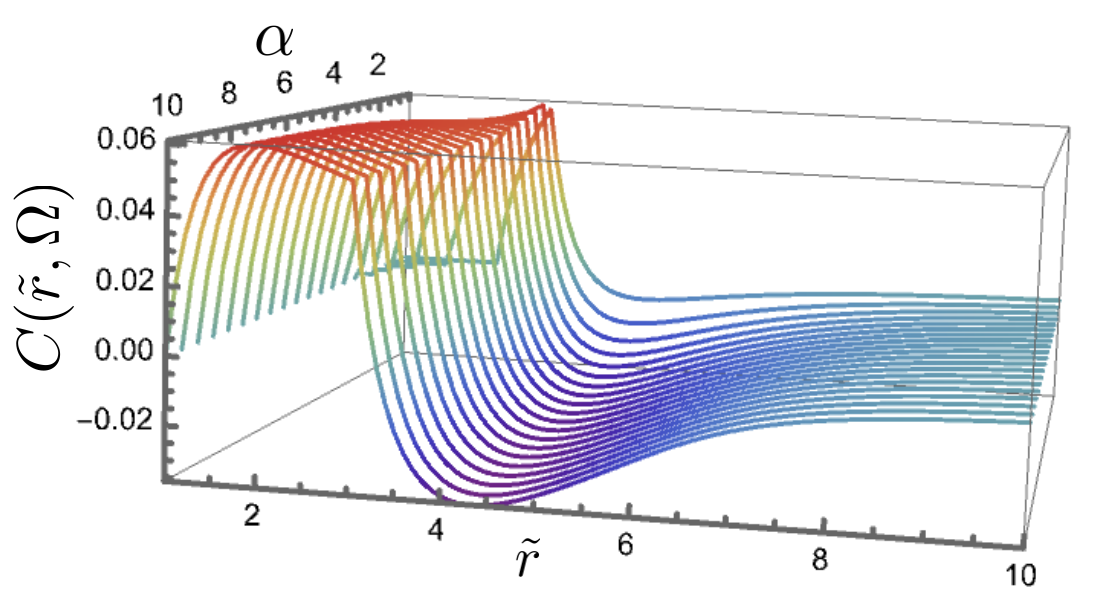}
\vspace*{-.5cm}\caption{Line plot of the correction term to  the Master equation  as function of
$\tilde{r},$ for various values of power $\alpha,$ for $\gamma=0$. } 
    \label{corr0}
\end{figure}

\begin{figure}
    \includegraphics[scale=0.23]{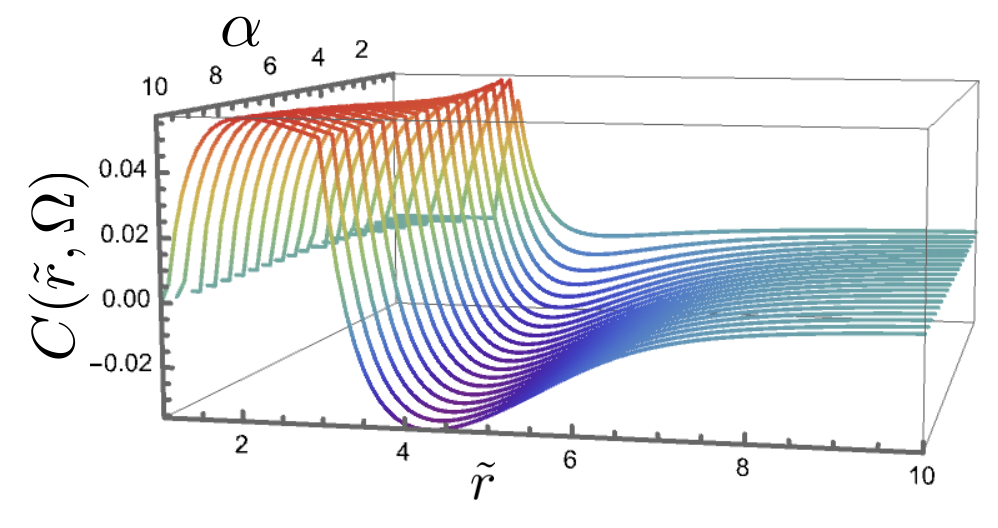}
\vspace*{-.5cm}\caption{ Line plot of the correction term to  the Master equation as function of
$\tilde{r},$ for various values of power $\alpha,$ for $\gamma=1$.} 
    \label{corr1}
\end{figure}

 Next, 
 we
   explore whether there are corrections to the  SDRG fixed point distribution
   Eq. (\ref{sdrgpr}), when  solving  the Master equation 
 with renormalized long range coupling Eq. (\ref{mlfinalc}), including the correction 
  due to the renormalization of distances, provided by the function 
 $ C(\tilde{r},\Omega),$  Eq. (\ref{c}). 
 Inserting the SDRG distribution Eq. (\ref{sdrgpr}) into Eq. (\ref{c}),
we plot the function  $ C(\tilde{r},\Omega)$ 
as function of
$\tilde{r},$ for various values of power $\alpha,$ for anisotropy  $\gamma=0$
in Fig.  \ref{corr0} and for $\gamma=1$ in  Fig.  \ref{corr1}.
We   see that  for both values of $\gamma$ it vanishes exactly at $\tilde{r}=\rho,$   $C(\rho,\Omega)=0,$ 
 increases with $\tilde{r}$ to a 
maximal  value close to $ r \approx  3\rho,$
decaying for  $r> 3\rho$ to negative values 
and converging to zero for  $r \gg 3\rho$. The magnitude of that correction  is slightly 
smaller for $\gamma=1,$ than for $\gamma=0.$ 
Remarkably, the integral over all distances, 
$K=\int_{\rho}^{\infty} d \tilde{r}  C(\tilde{r},\Omega) =0,$
is vanishing exactly for all 
 $\alpha.$ 
  
\begin{figure}
    \includegraphics[scale=0.23]{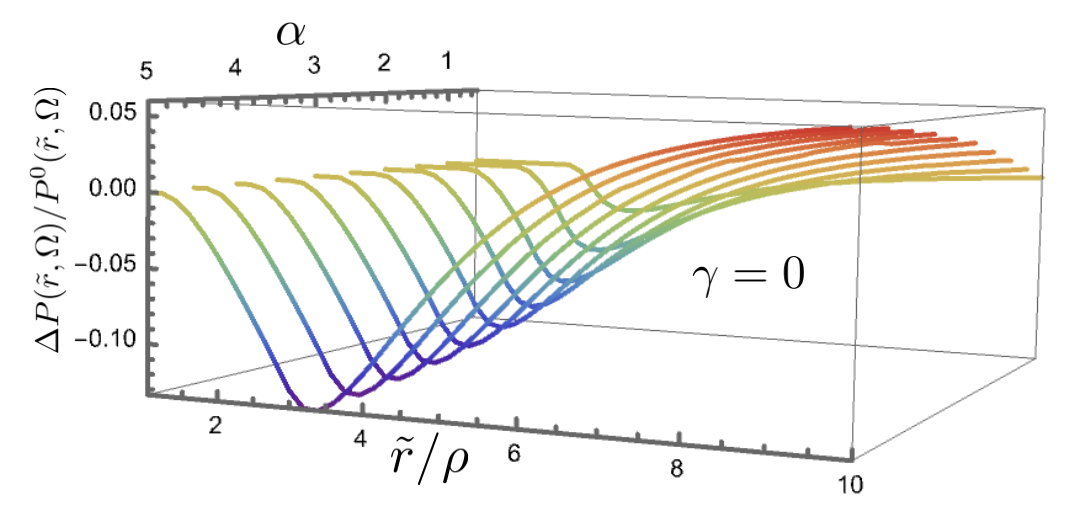}
\vspace*{-.5cm}\caption{  The ratio of 
  the correction to the distribution function 
  due to the renormalization  and the 
   SDRG distribution 
  Eq. (\ref{sdrgpr})
   as function of distance 
$\tilde{r}$ and power  $\alpha$ for  $\gamma=0$.} 
    \label{pdfg0}
\end{figure}

 By 
 inserting  the SDRG distribution Eq. (\ref{sdrgpr}) into Eq. (\ref{c}),
    the Master equation Eq. (\ref{mlfinal}) becomes a
    linear, 1st order inhomogeneous  differential equation.
Its solution is given by,
\begin{eqnarray} \label{prc}
 P(\tilde{r},\Omega) =  P^0(\tilde{r},\Omega) 
\left( 1-\int_1^{\tilde{r}/\rho} dx x^{1/2} c(x) \right),
\end{eqnarray}
where we defined the function 
\begin{eqnarray}
&&c(x) = 
 \frac{1}{4} \int_1^{\infty} d t_L \int_1^{\infty} d t_L (t_L t_R)^{-3/2} \times 
\nonumber \\
&&
(\delta ( x-\tilde{f} (t_L,t_R))- (\delta ( x- (t_L+t_R+1)),
\end{eqnarray}
where $\tilde{f}(t_L=R_L/\rho, t_R=R_R/\rho ) = f(R_L,R_R,\rho)/\rho,$
with $f$ given by Eq. (\ref{rgelr}).
The pdf  is normalized, $\int_{\rho}^\infty dr  P(r,\Omega)  =1,$
and we find  that $P(\rho,\Omega)= 1/(2 \rho)$, the same value as for the SDRG distribution.
\begin{figure}
    \includegraphics[scale=0.23]{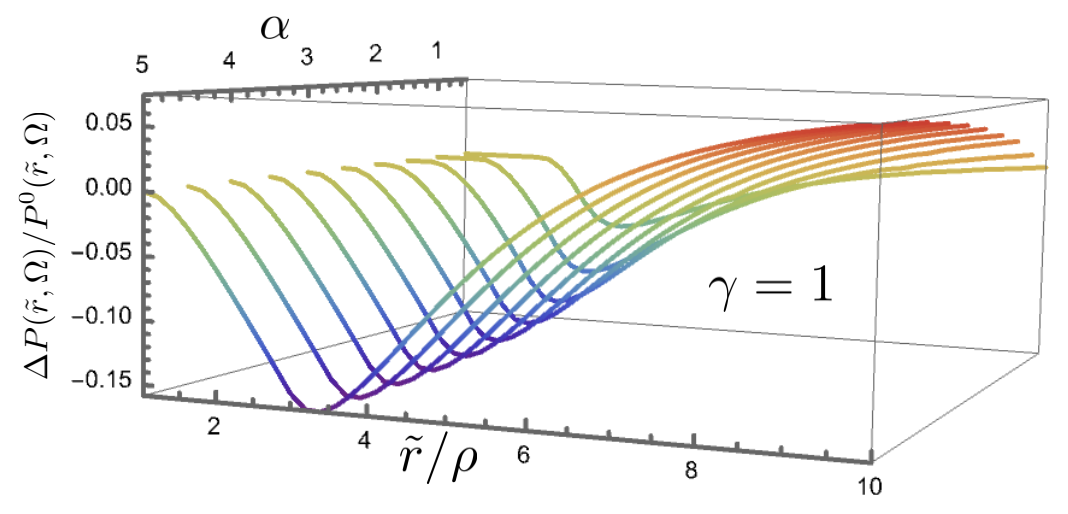}
\vspace*{-.5cm}\caption{  The ratio of 
  the correction to the distribution function  due to the renormalization and the  SDRG distribution 
  Eq. (\ref{sdrgpr})
   as function of distance 
$\tilde{r}$ and power  $\alpha$ for $\gamma =1$.} 
    \label{pdfg1}
\end{figure}  
In Fig. \ref{pdfg0} we plot 
 the ratio of 
  the deviation 
  of  the distribution function $\Delta P(\tilde{r},\Omega) =P(\tilde{r},\Omega)-P^0(\tilde{r},\Omega)$ 
  due to the renormalization 
  and the SDRG distribution  Eq. (\ref{sdrgpr})
 as function of distance 
$\tilde{r}$ and power  $\alpha$  
for
 $\gamma=0.$ 
and in  Fig. \ref{pdfg1} for  $\gamma=1$. 
We see that the correction due to the renormalization is negative and 
largest at $\tilde{r} \approx 3 \rho.$ 
It changes sign for larger distances  $\tilde{r}$ and approaches zero for  $\tilde{r} \gg 1.$
Its magnitude 
 increases with increasing $\alpha,$ but saturates for $\alpha \gg 1.$ It
 is slightly larger for $\gamma=0$  than for $\gamma=1$.  
  Since we obtained that solution iteratively by inserting the SDRG distribution 
  Eq. (\ref{sdrgpr}) into the correction term of the Master equation Eq.  (\ref{c}),
  we 
 insert that solution Eq. (\ref{prc}) back into the Master equation Eq.  (\ref{mlfinalc})
  with Eq.  (\ref{c})
 and find that the remaining terms are indeed  small,  of order $C^2,$
 and that these vanish for $\tilde{r} \rightarrow \rho.$ 
 We can therefore conclude that, while we  find  small corrections to the SDRG  distribution function at  distances  $\tilde{r} > \rho.$, and thereby to the distribution  of 
couplings, we find no correction to its value at $J=\Omega,$
\begin{equation} \label{pof}
P(\Omega,\Omega)= \frac{1}{2 \alpha \Omega},
\end{equation}
which  is also  found to be independent of  anisotropy $\gamma.$ 

\section{
Thermodynamic and Dynamic Properties}
\begin{figure}
    \includegraphics[scale=0.3]{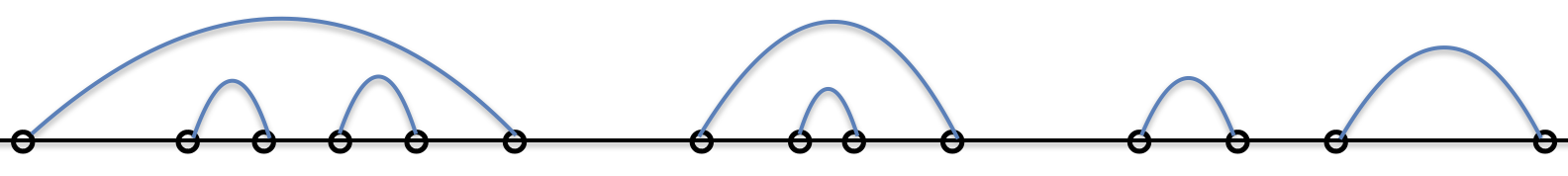}
\vspace*{-.3cm}\caption{Typical random singlet state. The blue lines 
 connect pairs of spins which formed a singlet state.} 
 \vspace*{-.5cm}
    \label{rss}
\end{figure}

Having 
derived  the   distribution function of 
 couplings for  the ground state,  a random singlet state as    sketched in Fig. \ref{rss},  allows us  to derive thermodynamic and  dynamic properties of bond  disordered spin chains, Eq. (\ref{H}), as summarized in the following. 

 The {\it magnetic susceptibility}  is given by
   $\chi(T) = n_{\rm FM} (T)/T,$ where the  density of free moments  
   at temperature $T,$     $n_{\rm FM} (T)$ is 
    governed by  the differential equation 
    \begin{eqnarray} \label{nfm}
 \frac{d n_{\rm FM}(\Omega)}{d \Omega} = 
2 P(J= \Omega, \Omega ) n_{\rm FM} (\Omega).
      \end{eqnarray} 
      Insertion of Eq. (\ref{pof}), performing the integration from $\Omega$
       to $\Omega_0,$
      yields  
        \begin{eqnarray} \label{nf}
       n_{\rm FM}(\Omega) = n_0 (\Omega/\Omega_0)^{1/\alpha}
        \end{eqnarray} 
        and  thereby  
  \begin{eqnarray} \label{chim}
\chi(T) = n_0  (\frac{T}{\Omega_0})^{\frac{1}{\alpha}} \frac{1}{T},
      \end{eqnarray} 
      independent of anisotropy $\gamma.$
     Thus, the 
      magnetic susceptibility diverges
       with   power
        $ 1-1/\alpha >0,$
        for $\alpha > \alpha^* =1.$
    It vanishes at $T=0K$ for  $\alpha < 1,$
       indicating  the presence of a pseudo gap in the density of states.

{\it Distribution of Singlet Lengths.—}
 The distribution of distances  $l$ between spins,
 bound into singlets in the random singlet state,
 $P_s(l),$ is determined  by  
 \begin{equation}  \label{psl}
 P_s(l) = c_s \frac{n_{FM}(\Omega)}{n_0} P(\Omega,\Omega)|_{\Omega = \Omega_0 l^{-\alpha}}
 | \frac{ d}{d l}  \Omega_0 l^{-\alpha}|,
 \end{equation}
 where $c_s$ is a normalization constant. 
Inserting Eqs. (\ref{pof},\ref{nf}) we find
 \begin{equation} \label{psl2}
P_s(l) \sim  n_0 (l n_0)^{-2},
 \end{equation}
  as previously derived  for the XX-Model \cite{Mohdeb2020}. Here we 
 find that it is valid for any anisotropy $0 \le \gamma \le 1.$
 
{\it Entanglement Entropy.—}
The entanglement entropy of a subsystem 
of length $n$
with the rest of the chain is for a
specific random singlet state,
as the one  shown in Fig. \ref{rss}, 
given by  $S_n = M \ln 2,$ where $M$ is the number of singlets crossing the partition of the subsystem. 
  The average entanglement entropy is 
  thereby given by
  $\langle S_n \rangle = 
  \langle M \rangle  \ln 2.$
 The average number of singlets 
 crossing  the partition of the subsystem
 can be derived 
from the distribution of singlet lengths $P_s(l),$
with the leading term given by\cite{refael,hoyos,Mohdeb2020} 
$S_n  \sim \frac{1}{2} \ln 2    \int_{l_0}^n dl ~ l P_s(l)$. 
This yields with  Eq. (\ref{psl})  $P_s(l) \sim l^{-2}$
 the logarithmic growth of entanglement entropy with 
 subsystem length $n,$
 \begin{equation}
S_n \sim \frac{1}{6} \ln 2 \ln n, 
 \end{equation}
for all $\alpha$ and anisotropy $\gamma.$
This  has  the  functional form of 
the entanglement entropy of 
critical quantum spin chains\cite{calabrese}
with effective central charge  $\bar{c} = \ln 2$\cite{refael},
 as was found previously in Ref. \cite{Mohdeb2020} for the XX-model 
 $\gamma =0,$  confirmed by numerical exact diagonalization.
  Here, we derived   it to be valid  for any anisotropy $0 \le \gamma \le 1$.

{\it
Entanglement Entropy Growth After a  Quantum Quench.—
}
  Preparing the system  in  an unentangled
  state, such as a  Néel state
$|\psi_0\rangle=|\uparrow\downarrow\uparrow\downarrow\uparrow...\rangle$,
the entanglement dynamics
 can be 
monitored by the  time dependent
entanglement entropy of a subsystem with the rest of the system $S(t)$. 
When   entanglement is generated 
by   
singlet or  entangled  triplet state  across the partition,
the entanglement entropy
at time $t$ after the global  quench
is proportional to the number of
such pairs  formed 
at RG-scale $\Omega \sim 1/t$\cite{Vosk2013}.
 Neglecting the history of
 previously formed pairs, 
 the number of newly formed pairs at RG scale
 $\Omega,$
 $n_{\Omega}$ 
is
$   dn_{\Omega}= P(J=\Omega,\Omega) d\Omega$ \cite{refael}.
Substituting   Eq. (\ref{pof}) we find 
$n_{\Omega} =  1/(2\alpha) \ln(\Omega)$.
Inserting  $\Omega\sim 1/t$ the entanglement entropy increases  with time as
 \begin{equation} \label{st}
    S(t) = S_p \frac{1}{2\alpha} \ln(t),
 \end{equation}
with the time-averaged  contribution of  pairs of spins $ S_p= 2\ln2- 1,$ when the initial state is a N\'eel state\cite{Vosk2013}. Then,  only singlet and  entangled triplet states are populated in the RSRG-t flow,   contributing equally.  This
  coincides  with 
  the result found in Ref. 
  \cite{Mohdeb2023}.
For the nearest neighbor XX spin chain with random bonds the growth  after a global quench is slower, $S(t) \sim\ln(\ln(t))$\cite{Vosk2013}.
This derivation neglects the 
 history of
 previously formed pairs. 
 In   a more accurate derivation, we  
  need to consider  
   that triplet states
    are renormalized differently than singlet states\cite{Mohdeb2022}, as can be done using 
our real space representation of the SDRG method, as will be presented in a subsequent 
publication\cite{Kettemann20252}.

\section{Conclusions}
 By implementing a real space representation of the 
  strong disorder renormalization group 
    we confirm   that bond 
 disordered antiferromagnetic  long range coupled quantum spin chains 
 are governed by the 
      strong disorder fixed point distribution
      Eq.  (\ref{pj}), with small corrections 
     depending
      on power exponent $\alpha$ and anisotropy $\gamma.$
  The low temperature 
      magnetic susceptibility is found to diverge
       with an  anomalous power law.
The distribution of singlet lengths  $l$  in the ground state  
 decays  as $l^{-2}$ and the   entanglement entropy of a subsystem of length $n$
 increases  logarithmically for all $\alpha$. The entanglement entropy 
   increases after a global quench logarithmically according to Eq. (\ref{st})
  with a prefactor decaying with increasing power $\alpha.$
Having obtained analytical  results for any anisotropy $\gamma,$ 
 with strong indications for the robustness of the SDRG fixed point in the asymptotic 
  low RG scale limit, 
  numerical studies, in particular finite size scaling 
  of numerical exact diagonalization   results are called for, as were done 
   in the XX limit in Ref. \cite{Mohdeb2020} and for a long range hopping model  of free  Fermions in 
   Ref. \cite{Juhasz2022}, where correction to the SDRG result for the  averaged entanglement entropy
    were found. We also stress that the SDRG procedure has to be understood as a statistical concept, where corrections to the random singlet state should can govern physical properties. F.e. typical  
 ensemble averages can be governed by corrections to RG, as already found for the short ranged model in Ref. \cite{fisher94} and for the long range XX model in Ref. \cite{Mohdeb2020}.

Having  established the   real space representation  of the 
  strong disorder renormalization group as a tool to derive new results on 
thermodynamic and dynamic properties of   disordered 
long range antiferromagnetically  coupled quantum  spin chains
 may   pave a new route to   study  
  long range coupled 
   disordered spin systems 
   also at finite temperature\cite{Kettemann20252} and 
   in higher dimensions. 

{\it Acknowledgments.-}
I acknowledge stimulating discussions with A. Ustyuzhanin. 
 I thank J. Vahedi for critical reading of the manuscript and feedback.

\section*{APPENDIX A:  Jordan Wigner Transformation.}
    It is insightful to use     the Jordan-Wigner transformation which maps    the spin chain  Eq. (\ref{H}) onto
    the Hamiltonian of  interacting fermions given by
    \begin{eqnarray}\label{jw}
       H &=& \sum_{i<j}  J^x_{i,j} \left( c_i^+ c_j e^{i \pi \hat{n}_{ij}} + c_j^+ c_i e^{-i \pi \hat{n}_{ij}}  \right) 
       \nonumber \\ 
      &+&  \sum_{i<j} J^z_{ ij}  (\hat{n}_{i} - \frac{1}{2}) (\hat{n}_{j} - \frac{1}{2}),
       \end{eqnarray}
       where the operator $\hat{n}_{ij} =\sum_{i<n<j} c_n^+ c_n $ counts how  many fermions 
       are encountered while hopping between the sites $i$ and $j$
        and $\hat{n}_{i} = c_i^+ c_i$ is the density operator of fermions at site $i$.
      For  $J^z_{ij}=0$ and
      nearest neighbor hopping this is  the 
       Hamiltonian of  noninteracting  fermions with random hopping,
        which is known to  show the Dyson anomaly:
        the eigenfunctions in the center of the band
         decay spatially with a stretched exponential,
           $\psi(x) \sim \exp (- \sqrt{x/l_0}]), $ where $l_0$ is a small length scale\cite{dyson}. 
           Away from the band center the eigenfunctions decay exponentially with localization length $\xi$, which diverges at the band center as 
            $\xi \sim - \ln |\epsilon|.$
          The density of states is singular at half filling, 
           $\rho(\epsilon) = |\epsilon|^{-1} \ln |\epsilon|^{-3}$\cite{wegner}.        
       With long  range hopping,  $J^x_{ij} \neq 0$
    the interaction between the fermions 
     manifests itself 
      through the  dynamic phases $\pi \hat{n}_{ij}$ in the hopping amplitudes.
      Direct interaction between the fermions occurs for any   $J^z_{ ij} \neq 0$.

      \section*{APPENDIX B:  Derivation of Master Equation with Long Range Couplings.}

In the representation of distances  $r$ 
we need to derive 
the Master equation for
 the distribution function
 $P(\tilde{r},\Omega),$
which is defined to be the pdf for the distances between adjacent spins at RG scale $\Omega$. 
 Thus, for the chain with  open boundary conditions  it is the distribution of 
  the $N-1$ nearest neighbor distances in the chain.
 Note that the distances between non adjacent spins 
 in the chain are  functions of these 
 $N-1$ adjacent distances and should not be counted in,  
  when calculating  $P(\tilde{r},\Omega)$. 
When the singlet between 
spin pair  $(i,j)$ is formed
 at RG scale $\Omega = J_{ij},$ corresponding to  a distance 
   $\rho = (\Omega_0/\Omega)^{1/\alpha},$
 the two  bonds between the two adjacent  spin pairs
 $(l,i)$ and $(j,m)$  shown in Fig. \ref{RG} 
  with distances  
  $r_{l,i}=R_L$  and $r_{i,m}=R_R$
 are taken away.
    The   bare coupling  $J_{lm}$ is    then 
    renormalized into the  coupling $\tilde{J}_{lm}$.
     In the representation of distances 
      this corresponds to creating an adjacent bond with   renormalized distance $\tilde{r}_{lm}=\tilde{r}$  
replacing their previous distance 
${r}_{lm}={r},$ as indicated in  Fig. \ref{RG}.
For other adjacent 
    spin pairs like  $l',m'$ ( or likewise $l'',m''$) in Fig. \ref{RG},  where both 
    spins  $l',m'$ are located on 
    the same side
    of the singlet $(i,j),$ 
 their bare coupling is  
    renormalized into the  coupling $\tilde{J}_{l'm'}$.
     In terms of the representation of distances 
      this corresponds to the  creation of an adjacent bond with  renormalized distance $\tilde{r}_{l'm'}=\tilde{r}$,
replacing their previous distance 
${r}_{l'm}'=\tilde{r}.$  
However, for such pairs the renormalization is small, of the order of 
$(\rho/R)^{2\alpha+2},$ where $R$ is the distance between the pair $l',m'$
and the removed pair $i,j.$  

 Thus, the removal 
   of  pair $(i,j)$ leads to  
      the renormalization of 
  all remaining spin positions  ${\bf r}_l$, as sketched in Fig. \ref{RG}, 
so that  distances
  between  spins at  sites $l,m$  on different sides of $i,j$ are shortened, 
  while distances between spins  on the same side of  $i,j$
  remain unchanged. 
It is remarkable that while    all couplings become  renormalized, 
   this  can be fully accounted for by 
 the renormalization of the distance between the single  pair of spins which become neighbors after the renormalization step, $r_{lm}$, in the notation used in Fig. \ref{RG}.
  All other renormalied distances follow from that by simple addition of distances. 
  Let us take for example the pair of  spins located on different sides of the removed pair $(m',l'')$  shown  in Fig. \ref{RG}. Then,  the fact that its renormalized distance follows from the 
    renormalization of $r_{lm}$ into $\tilde{r}_{lm}$
can be  cast into the formula 
 $ \tilde{r}_{m'\prime l''\prime} = \tilde{r}_{lm}   - r_{m'\prime l}  - r_{ml''\prime}$,  
where 
 $\tilde{r}_{m'\prime l''\prime }$ 
is the   renormalized distance between the pair  of spins at sites to the left and right of the removed pair $(ij)$, which become not neighbours after the RG-step. 
That equality can be proven as follows. 
By plotting the renormalized distance as function of $R_L$ and $R_R$  in   Fig.  \ref{ff}.
 we established 
 the inequality $r_{lm} - 2 \rho    < \tilde{\rho}_{lm}  < r_{lm}$.
That inequality  applies also to the renormalization of distances between  pair  of spins at sites   to the left and right of the removed pair, like   $(m',l'')$  shown  in Fig. \ref{RG}, which become not
neighbors after the RG-step:
           $r_{m'\prime l''\prime } - 2 \rho    < \tilde{\rho}_{m'\prime l''\prime }  < r_{m'\prime l''\prime }$.
Inserting the relation   $r_{m’\prime l''\prime }  = r_{lm}  +r_{m'\prime l} +r_{ml’'\prime } $    (which follows from  the  definition of  distances), we find 
           $r_{lm} - 2 \rho   < \tilde{r}_{m’\prime l''\prime }  - r_{m'\prime l}  - r_{ml’'\prime }   < r_{lm}$
which is consistent with the equality we wanted to prove,
 which proves our claim, as  implied in the  sketch of the RG step  in Fig. 1. 
Therefore, only 
the renormalization of the distance of adjacent spins $(l,m)$ 
needs to be taken into account in the 
 derivation of  the Master equation for the distribution function  $P(\tilde{r},\Omega).$
 
 Thus, 
 we obtain  for the distribution function of $\tilde{r}$ at the
 reduced energy scale $\Omega-d\Omega$ 
\begin{eqnarray} \label{pr2}
 &&P(\tilde{r},\Omega-d\Omega) =
 \left( P((\tilde{r},\Omega)  +  d\Omega P(\Omega,\Omega)   \int_{\rho}^{\infty} d r \int_{\rho}^{\infty} dR_L 
 \right.
\nonumber \\
 && 
 \int_{\rho}^{\infty}
 dR_R  P(R_L,\Omega) P(R_R,\Omega)  \delta(r-R_L-\rho-R_R)   \times  \nonumber \\
 &&  
 (\delta (\tilde{r}- f(r,R_L,R_R,\rho))-\delta (\tilde{r}- r)-\delta (\tilde{r}- R_L)
   \nonumber \\
 && \left.
 -\delta (\tilde{r}- R_R))
\right)
 \frac{1}{1-3 d\Omega  P(\Omega,\Omega)}.
      \end{eqnarray}  
      In  the renormalization  term on the right side of Eq. (\ref{pr2}), 
      we defined the bare distances,  $r=r_{lm},$ 
      $R_L =r_{li},$ and  $R_R =r_{jm},$
     with the constraint on  the bare distance $r_{lm}=r = R_L +R_R +\rho,$ as implemented 
      by a delta-function.
 The  delta-functions in the bracket account for the fact 
  that 
 one adjacent edge is created 
  between $l$ and $m$ 
  with distance $\tilde{r},$ removing the one with distance $r,$ and removing the two adjacent bonds with distances 
   $R_L$ and $R_R.$  
     The
      renormalized  
      distance $\tilde{r} $  is according to Eq. (\ref{reff}) given by 
      \begin{eqnarray} \label{rgelr}
      &&  f(r,R_L,R_R,\rho)=r ( 1+ \frac{1}{1+\gamma} (\frac{r \rho}{R_L R_R})^{\alpha} \times  
      \nonumber \\ 
     && (1-\frac{1}{(1+\rho/R_L)^{\alpha}})
      (1-\frac{1}{(1+\rho/R_R)^{\alpha}}) )^{-1/\alpha}.
      \end{eqnarray}
 The last factor on the right side of Eq. (\ref{pr2}) is needed for normalization of the pdf, since 
 in total 3 edges are taken away. The proper normalization can be checked, by integrating both sides
  of Eq. (\ref{pr2}) over $\tilde{r}$ from $\rho$ to infinity, 
  Taylor expanding in $d \Omega$
 and  using the normalization 
  condition $\int_{\rho}^{\infty} d \tilde{r} P(\tilde{r},\Omega) =1$
 and  $\rho = (\Omega_0/\Omega)^{1/\alpha}.$
 
To be able to cancel 
 the normalization factor in Eq. (\ref{pr2}) we need to substract and add another  term 
 $d\Omega P(\Omega,\Omega) P(\tilde{r},\Omega)$.
In the limit $d\Omega \rightarrow 0$
we  thereby find 
the Master equation as  
\begin{eqnarray} \label{mlfinal}
 &&- \frac{d}{d \Omega} 
 P(\tilde{r},\Omega) =
P(\Omega,\Omega) \left( P(\tilde{r},\Omega) + \int_{\rho}^{\infty}
 dR_L \int_{\rho}^{\infty}
 dR_R  \right.
  \nonumber  \\
 &&    \left. P(R_L,\Omega) P(R_R,\Omega)
 ( \delta (\tilde{r}- f(r=R_L+\rho+R_R,R_L,R_R,\rho))
  \right.
  \nonumber  \\
 &&    \left. 
 -  \delta (\tilde{r}-(R_L+\rho+R_R)) \right).
      \end{eqnarray}


\begin{thebibliography}*
      
 
 \bibitem{anderson58} 
P. W. Anderson, {Absence of diffusion in certain random lattices}, Phys. Rev. {\bf 109}, 1492 (1958).

\bibitem{loehneysen}  H. v. L\"ohneysen,
Disorder, electron-electron interactions and the metal-insulator transition in heavily doped Si:P,
 Adv. in Solid State Phys. {\bf 40}, 143 (2000).


\bibitem{Bhatt2021} 
R.N. Bhatt and S. Kettemann.  Special Issue Localisation 2020: Editorial Summary, 
 Annals of Physics  {\bf 435},
168664 (2021). 

\bibitem{Kettemann2023} {  S. Kettemann,
Towards a Comprehensive Theory of Metal-Insulator Transitions in Doped Semiconductors,
 Special Issue in memory of K. B. Efetov, 
Annals of Physics
 456, 169306 (2023). }

  \bibitem{Kettemann2024} 
 S. Kettemann, 
 Competition between Kondo Effect and RKKY Coupling,
Lecture Notes of the Autumn School on Correlated Electrons 2024, Correlations and Phase Transitions, Band/Volume 14 edited by Eva Pavarini and Erik Koch,Verlag des Forschungszentrums J\"ulich (2024).


   \bibitem{reviewtls}
D. Salvino, S. Rogge, B. Tigner, D. Osheroff,
{  Low Temperature ac Dielectric Response of Glasses to High dc Electric Fields},
Phys. Rev. Lett. {\bf  73}, 286 (1994).


\bibitem{Yu} C. C. Yu and A. J. Leggett,
{ Low temperature properties of amorphous materials: Through a glass darkly}, 
 Commun. Condens. Mat. Phys.
{\bf 14}, 231 (1988). 


 \bibitem{Bilmes2020}
A. Bilmes,  et al,
A. Megrant, P. Klimov, G. Weiss, J. M. Martinis, A. V.  Ustinov,  J. Lisenfeld,  
{Resolving the positions of defects in superconducting quantum bits}, 
Scientific reports {\bf 10}, 1-6 (2020).




 \bibitem{Lukin2022}   B. L. Dwyer, L. V. H. Rodgers, E. K. Urbach, D. Bluvstein, S. Sangtawesin, H. Zhou, Y. Nassab, M. Fitzpatrick, Z. Yuan, K. De Greve, E. L. Peterson, H. Knowles, T. Sumarac, 
    J.-P. Chou, A. Gali, V.V. Dobrovitski, M. D. Lukin and N. P. de Leon,
    { Probing spin dynamics on diamond surfaces using a single quantum sensor}, 
PRX Quantum {\bf 3}, 040328  (2022). 

  \bibitem{Davis2022} E. J. Davis,  B. Ye, F. Machado, S. A. Meynell, W. Wu, T. Mittiga, W. Schenken, M. Joos, B. Kobrin, Y. Lyu, Z. Wang, D. Bluvstein, S. Choi, C. Zu, A. C. Bleszynski Jayich, N. Y. Yao, 
 { 
Probing many-body dynamics in a two dimensional dipolar spin ensemble}, 
 Nature Physics, https://doi.org/10.1038/s41567-023-01944-5 (2023). 
   
   \bibitem{douglas} 
J. S. Douglas, H. Habibian, A. V. Gorshkov, H. J. Kimble, D. E. Chang, 
{ Atom induced cavities and tunable long-range interactions between atoms trapped near photonic crystals},
	Nature Photonics 9, 326-331 (2015).
 
\bibitem{grass} T.  {Gra\ss }   
and M. Lewenstein, 
{ Trapped-ion quantum simulation of tunable-range Heisenberg chains},
 EPJ Quantum Technology 1:8 (2014).
   
 \bibitem{Signoles2021} A. Signoles, T. Franz, R. F. Alves, M. G\"arttner, S. Whitlock, G. Z\"urn, and M. Weidem\"uller, 
 { 
Glassy dynamics in a disordered Heisenberg quantum spin system}, 
 Phys. Rev. {\bf X 11},
011011 (2021).



 \bibitem{Brow2020} A. Browaeys and T. Lahaye, 
 { Many-body physics with individually controlled Rydberg atoms}, 
 Nature Physics {\bf 16}, 132 (2020).

\bibitem{Franz2022} 
T. Franz, S. Geier, C. Hainaut, A. Signoles, N. Thaicharoen, A. Tebben, A. Salzinger, A. Braemer, M. G\"arttner, G. Z\"urn, and M. Weidem\"uller, 
 Emergent pair localization in a many-body quantum spin system,
 arXiv.2207.14216 (2022).


\bibitem{Islam2013} R. Islam, C. Senko, W. C. Campbell, S. Korenblit, J. Smith,
A. Lee, E. E. Edwards, C.-C. J. Wang, J. K. Freericks, and C.
Monroe, 
Emergence and Frustration of Magnetic Order with Variable-Range Interactions in a Trapped Ion Quantum Simulato,
Science {\bf 340}, 583 (2013).

\bibitem{Richerme2014} 
 P. Richerme, Z.-X. Gong, A. Lee, C. Senko, J. Smith, M. Foss-
Feig, S. Michalakis, A. V. Gorshkov, and C. Monroe,
Non-local propagation of correlations in quantum systems with long-range interactions,
Nature
(London) 511, 198 (2014).

\bibitem{Jurc2014} 
 P. Jurcevic, B. P. Lanyon, P. Hauke, C. Hempel, P. Zoller, R.
Blatt, and C. F. Roos,
Quasiparticle engineering and entanglement propagation in a quantum many-body system,
Nature (London) 511, 202 (2014).

       
\bibitem{bhattlee81} R. N. Bhatt and P. A. Lee, 
{ A scaling method for low temperature behavior of random antiferromagnetic systems},
 Journal of Applied Physics {\bf 52}, 1703-1707 (1981).
    
\bibitem{bhattlee82}   R. N. Bhatt and P. A. Lee,
{ Scaling studies of highly disordered spin-1/2 antiferromagnetic systems}, 
 Phys. Rev. Lett. {\bf 48}, 344(1982).
 
 \bibitem{fisher94}  
 D. S. Fisher,
{ Random antiferromagnetic quantum spin chains},
  Phys. Rev. B {\bf 50}, 3799 (1994).
  
  

\bibitem{fisher95}  
 D. S. Fisher,
{ Critical behavior of random transverse-field Ising spin chains
},
  Phys. Rev. B {\bf 51}, 6411 (1995).
    
 \bibitem{monthus}
F. Igloi and C. Monthus, 
{ Strong disorder RG approach of random systems},
Phys. Rep.  {\bf 412}, 277 (2005). 

  
 \bibitem{Igloi2018} F. Igl\'oi and  C. Monthus, 
{ Strong disorder RG approach - a short review of recent developments },
Eur. Phys. J. {\bf B  91}, 290
 (2018).


  \bibitem{Moure2015} 
N. Moure, S. Haas, S. Kettemann,
 {  Many Body Localization Transition in Random Quantum Spin Chains with Long Range Interactions},
 Europhys. Lett.  {\bf 111}, 27003 (2015). 


 
 \bibitem{Moure2018}  {
N. Moure, Hyun-Yong Lee, S. Haas, R. N. Bhatt, S. Kettemann, 
{ Disordered Quantum SCs with Long-Range Antiferromagnetic Interactions},  Phys. Rev. {\bf B 97}, 014206  (2018). }
 
\bibitem{Mohdeb2020}  {
  Y. Mohdeb, J. Vahedi, N.  Moure,  A. Roshani, H.Y. Lee,  R. Bhatt, S. Haas, S. Kettemann, 
  Entanglement Properties of Disordered Quantum SCs with Long-Range Antiferromagnetic Interactions,  Phys. Rev.  {\bf B 102}, 214201   
   (2020). }



\bibitem{Juhasz2014} 		
R. Juh\'asz, I. A.  Kov\'acs, F.  Igl\'oi, 
{ Random transverse-field Ising chain with long-range interactions}, 
Europhys. Lett.,  {\bf 107}, 47008 (2014).

\bibitem{Juhasz2016} 
I. A.  Kov\'acs, R.  Juh\'asz,   F.  Igl\'oi, 
{ Long-range random transverse-field Ising model in three dimensions}, 
 Phys. Rev. B {\bf 93}, 184203 (2016).


 \bibitem{Pekker2014}  D. Pekker, G. Refael, E. Altman, E. Demler, and V. Oganesyan,
Phys. Rev. X 4, 011052 (2014).

\bibitem{Vosk2013}
R. Vosk and E. Altman, 
{ Many-Body Localization in One Dimension as a Dynamical Renormalization Group Fixed Point}, 
Phys. Rev. Lett. {\bf 110}, 067204 (2013).

\bibitem{Vosk2014} R. Vosk and E. Altman,
{ Dynamical quantum phase transitions in random spin chains}, 
 Phys. Rev. Lett. {\bf 112}, 217204 (2014).

\bibitem{Igloi2012} F. Igloi, Z. Szatm\'ari, and Y-C. Lin,
{ Entanglement entropy dynamics of disordered quantum spin chains}, 
 Phys. Rev. B {\bf 85}, 094417 (2012).

 
  \bibitem{Mohdeb2022} { Y. Mohdeb, J. Vahedi, S. Kettemann, Excited-Eigenstate Properties of  XX Spin Chains with Random Long-Range Interactions,
 Phys. Rev. B {\bf B  106}, 104201 (2022). }

    


     
 \bibitem{Mohdeb2023}
{ Y. Mohdeb, J. Vahedi, S. Haas, R. N. Bhatt, S. Kettemann,
{ Global Quench Dynamics and the Growth  of Entanglement Entropy in Disordered Spin Chains with Tunable Range Interactions},  Phys. Rev. B   108, L140203  (2023). 
 }

 


\bibitem{refael}
 G. Refael and J. E. Moore, 
 Entanglement Entropy of Random Quantum Critical Points in One Dimension,
 Phys. Rev. Lett. {\bf 93}, 260602 (2004). 

 \bibitem{hoyos}
 J. A. Hoyos, A. P. Vieira, N. Laflorencie, and E. Miranda, 
 Correlation amplitude and entanglement entropy in random spin chains,
 Phys.  Rev. {\bf  B 76}, 174425 (2007). 

 \bibitem{calabrese}
 P. Calabrese and J. Cardy,
Entanglement entropy and quantum field theory,
 J. Stat. Mech.: Theory Exp. P06002 (2004).

 \bibitem{Juhasz2022} R. Juhasz, Phys. Rev. {\bf  B. 105}, 014206 (2022).


\bibitem{Kettemann20252} S. Kettemann, unpublished (2025). 


   \bibitem{dyson} F. J. Dyson, 
 The dynamics of a disordered linear chain,
Phys. Rev. {\bf 92}, 1331(1953).

\bibitem{wegner} F. J. Wegner, in {\it Fifty Years of Anderson Localization},
World Scientific (2010);  arXiv:1003.0787.




  \end{thebibliography}
\end{document}